\documentclass[runningheads]{llncs}
\usepackage{amsfonts}
\usepackage{amscd}
\usepackage{bbm}
\usepackage{amsfonts,amssymb}
\usepackage{amsmath}
\usepackage[bookmarksopen]{hyperref}
\usepackage{color}
\usepackage{booktabs}
\usepackage{dsfont}

\usepackage{amsmath}
\usepackage{amscd}
\usepackage{amsfonts,amssymb}
\usepackage[bookmarksopen]{hyperref}
\usepackage{color}
\usepackage{dsfont}
\usepackage{graphicx}
\usepackage{epstopdf}
\usepackage{threeparttable}
\usepackage{subfigure}
\usepackage{lineno,hyperref}
\usepackage{indentfirst}

\usepackage{diagbox}
\usepackage{enumitem}
\usepackage{fancyhdr}

\begin{document}
	\title{Provably Secure Non-interactive Key Exchange Protocol for Group-Oriented Applications in Scenarios with Low-Quality Networks}

\author{Rui Zhang\inst{1,2} \and Lei Zhang\inst{1\thanks{Corresponding Author}}}

\institute{Software Engineering Institute, East China Normal University \and Computer Science and Artificial Intelligence Institute, Southwestern University of Finance and Economics}

\pagestyle{plain}

\maketitle

\begin{abstract}
Non-interactive key exchange (NIKE) enables two or multiple parties (just knowing the public system parameters and each other’s public key) to derive a (group) session key without the need for interaction. 
Recently, NIKE in multi-party settings has been attached importance. However, we note that most existing multi-party NIKE protocols, underlying costly cryptographic techniques (i.e., multilinear maps and indistinguishability obfuscation), lead to high computational costs once employed in practice. Therefore, it is a challenging task to achieve multi-party NIKE protocols by using more practical cryptographic primitives. 

\setlength{\parindent}{2em}
\par In this paper, we propose a secure and efficient NIKE protocol for secure communications in dynamic groups, whose construction only bases on bilinear maps. This protocol allows multiple parties to negotiate asymmetric group keys (a public group encryption key and each party’s decryption key) without any interaction among one another. Additionally, the protocol supports updating of group keys in an efficient and non-interactive way once any party outside a group or any group member joins or leaves the group.
Further, any party called a sender (even outside a group) intending to connect with some or all of group members called receivers in a group, just needs to generate a ciphertext with constant size under the public group encryption key, and only the group member who is the real receiver can decrypt the ciphertext to obtain the session key. We prove our protocol captures the correctness and indistinguishability of session key under $k$-Bilinear Diffie-Hellman exponent ($k$-BDHE) assumption. Efficiency evaluation shows the efficiency of our protocol. 	
\end{abstract}


\section{Introduction}
Non-interactive key exchange (NIKE) is a fundamental cryptographic primitive that enables two or multiple parties, who just know the public system parameters and each other's public key, to agree on a session key without any interaction among one another. Since NIKE eliminates the communication complexity during the session key establishment, it could be potentially applied into many real-world applications, especially those with limited bandwidth resources (i.e., wireless sensor networks, fog computing). For example, in wireless sensor networks, the battery power consumption is a prime concern. Using NIKE to establish a session key is helpful in minimising each sensor's energy cost of communication to a large extent. In addition, NIKE effectively prevents an adversary from interfering the key establishment process by the wireless radio, since no interaction is required. The multi-party NIKE scheme has broader application scenarios compared to the two/three-party one. For instance, it can be applied to scenarios with limited communication resources, such as mobile ad-hoc networks (MANETs) and wireless sensor networks (WSNs), to secure communication among a group of nodes. However, conventional multi-party NIKE schemes may still face the following challenges when deployed practical

Most distributed systems have the dynamic feature, such as changeable topology in MANETs, which necessitates frequent updates of the session key.
As for this, traditional multi-party NIKE tends to re-negotiate a session key by performing the scheme again. This might lead to unnecessary computational costs, especially when the there are minor changes of the group membership. 
Also, there exists an entity outside a group who wants to connect with some group members of a group. However, in traditional multi-party NIKE, the session key is only known by group participants, which implies the outsider cannot contact any group member unless it joins the group to derive a new session key. 
Moreover, in these distributed systems, the outsider is allowed to choose its preferred group members within a group for connection. For example, in MANETs, the device wants to choose those sensors that run stably with good performance to obtain some sensible data. 
It is obvious to see that traditional NIKE schemes don’t consider these demands in practical or satisfy some of these requirements but at the cost of sacrificing efficiency.

Most recently, based on asymmetric group key agreement (AGKA) \cite{wu2009asymmetric} and contributory broadcast encryption (CBE) (an extension of AGKA) \cite{wu2011bridging,zhang2019key,chen2021blockchain}, some one-round key exchange protocols in multi-party environments have been proposed \cite{zhang2015round,li2017sender}, which allow multiple parties to agree on the asymmetric group keys (e.g., a public group encryption key and each party's unique decryption key). Anyone can send an encrypted message (under a group encryption key) to a group since the key is public. This novel feature makes the protocols suitable for distributed applications where parties might live in different time zones. To eliminate round complexity, an innovative idea is to design non-interactive AGKA/CBE (NI-AGKA/NI-CBE) protocol. The original non-interactive AGKA protocol was introduced in \cite{wu2009asymmetric} based on the idea of trivial broadcast encryption. However, in this protocol, the computation cost for a sender and the size of the ciphertext both increase linearly with the number of receivers.
Using the idea in \cite{wu2009asymmetric}, one can get NI-CBE but it still has the same limitation as the original NI-AGKA. As so far, no CBE-based NIKE protocol with constant-size ciphertext has been proposed. 

\subsection{Related Work}
Over the years, NIKE has been well studied theoretically and practically. According as whether the number of users participating in key exchange is more than three, we categorize the existing NIKE protocols into two types: two-party/three-party NIKE protocol and multi-party NIKE protocol (where more than three users form a group). 
The original Diffie-Hellman key exchange protocol \cite{diffie1976multiuser} is known as the typical example of two-party NIKE. Later, some concrete two-party NIKE protocols were instantiated \cite{freire2013non,freire2014universally} based on different assumptions (e.g., factoring assumption and strong DH assumption). The first one-pass three-party key exchange protocol was proposed by Joux in \cite{joux2000one} which can be the basis of constructing three-party NIKE protocol.
Simultaneously, there has been emerged lots of research regarding the implementation of two/three-party NIKE protocols in practical applications \cite{boneh2003applications,ccapar2013signal}. Despite its efficient use in practice, the first type of NIKE has a limited application scope, only being applicable in two/three party communication scenarios.

The second type of NIKE aims to establish a session key shared by a group of users in a non-interactive way, hence it could be used to facilitate communications in group-oriented applications. Existing multi-party NIKE protocols are mostly constructed under multilinear maps (MMPs) and indistinguishability obfuscation (IO). For instance, in \cite{boneh2003applications}, Boneh et al first proposed the multiparty NIKE protocol underlying MMPs. In the sequel, several multi-party NIKE protocols based on MMPs were discussed\cite{boneh2003applications}, as many candidate MMPs were proposed \cite{garg2013candidate,coron2013practical,cheon2015cryptanalysis,gentry2015graph}. The first IO-based NIKE protocol in multiparty settings was proposed in \cite{rao2014adaptive}, which was further improved regarding security \cite{boneh2017multiparty,khurana2016multi}. However, we note that MMPs/IO itself isn't lightweight. This implies constructing MMPs/IO-based NIKE protocols could be costly in practice. Moreover, achieving a secure MMP is still an open problem as many candidate MMPs suffer from security issues \cite{hu2016cryptanalysis}. Obviously, multi-party NIKE protocols based on MMPs/IO are not practical enough in the real scenarios in terms of security and efficiency.

To define the security of two/three/multi-party NIKE protocols above, various security models have been proposed. Among them, the CKS model is known for the first security model in two-party NIKE settings \cite{cash2009twin}. Based on CKS model, Freire et.al, \cite{freire2013non} formalized various security models suiting for two-party NIKE based on different assumptions. The first security model for multi-party NIKE was introduced in \cite{boneh2020multiparty}. Based on this, we note that in multi-party settings, the basic security properties that a secure NIKE protocol should satisfy contain the consistency and the indistinguishability of the shared session key. The former means every participating party has to derive the same session key as other participating parties while the latter requires the session key is indistinguishable from a uniform random string in the view of any party who is not the participant. 
The indistinguishability of the session key further guarantees the confidentiality of a sent message. 



\subsection{Our Contribution}
Motivated by the observations above, we propose a multi-party NIKE protocol, called as non-interactive contributory broadcast encryption (NI-CBE) protocol, which is the first CBE-based NIKE protocol and specifically suitable for dynamic groups. The details of our protocol are shown as follows:
\begin{itemize}[leftmargin=*]
	\item Our NI-CBE protocol allows a group of users to negotiate a public group encryption key and each user's decryption key without requiring any interaction. Besides, our protocol suits for a group with dynamic membership, that is, any inside member/outside user is allowed to join/leave the group at any time but still without any interaction. Our protocol also supports any user (even outside a group) to connect with any group member in a group. Specifically, anyone can encrypt a message, e.g., a session key under this encryption key for some/all group members in a group since the key is publicly obtained. And only those group members selected by the sender can decrypt the ciphtertext by using their own decryption key to get the session key. We note that this process is completed without any communication costs. 
	\item The NI-CBE protocol is instantiated based on bilnear maps, whose security relies on the $k$-BDHE assumption. Specifically, a semi-static security model is designed. Based on the model, we prove our NI-CBE protocol captures the indistinguishability of session key, which further implies the session key remains confidential to anyone who is not selected by a sender to decrypt the session key. We note that the NI-CBE protocol guarantees the confidentiality of the session key even if an adversary obtains the decryption keys of all group members except those selected group members by the sender. Finally, efficiency evaluation shows the utility of our protocol. 
\end{itemize}

\section{Preliminaries}\label{Bilinear Maps}
\subsection{Bilinear Maps}
Our protocol is based on the bilinear maps. Let $\mathbb{G}_1$ and $\mathbb{G}_T$ be two multiplicative groups of
prime order $q$, and $g$ be a generator of $\mathbb{G}_1$. A map
$\hat{e}: \mathbb{G}_1\times \mathbb{G}_1 \rightarrow \mathbb{G}_T$
is called a bilinear map if it satisfies the following conditions:
\begin{itemize}
	\item Bilinearity: $\hat{e}(g^{\alpha}, g^{\beta}) = \hat{e}(g, g)^{\alpha\beta}$
	for all $\alpha, \beta\in \mathbb{Z}_q^*$.
	\item Non-degeneracy: There exist $a\in \mathbb{G}_1, b\in \mathbb{G}_2$ such that $\hat{e}(a, b)\neq 1$.
	\item Computability: For any $a, b\in \mathbb{G}_1$,
	$\hat{e}(a, b)$ can be calculated efficiently.
\end{itemize}

\subsection{Complexity Assumption}
The security of our protocol is reduced to the decision $k$-Bilnear Diffie-Hellman exponent (BDHE) assumption, which is first introduced in \cite{boneh2005hierarchical}.

\smallskip\noindent
\emph{Decision $k$-BDHE problem}: Given a bilnear map $\hat{e}: \mathbb{G}\times \mathbb{G} \rightarrow \mathbb{G}_T$. Let $g$ be a generator of $\mathbb{G}$, $Q=g^h$ for unknown $h \in \mathbb{Z}_q$, and $\mathbb{X}=\{X_i=g^{\theta^i}\}_{\{i = 1,2,...,k,k+2,..., 2k\}}$ for unknown $\theta \in \mathbb{Z}_q^*$. An algorithm $\mathcal{D}$ that outputs $b\in \{0,1\}$ has advantage
$\epsilon$ in solving the decision $k$-BDHE problem if
\begin{equation*}
	\begin{aligned}
		|\Pr[\mathcal{D}(g,Q,X,Z_0)=0]-  \Pr[\mathcal{D}(g,Q,X,Z_1)=0]|\geq \epsilon
	\end{aligned}
\end{equation*}
where $Z_0=\hat{e}(g^{\theta^{k+1}}, Q)$ and $Z_1\in \mathbb{G}_T$ randomly. The decision $k$-BDHE assumption holds in $\mathbb{G}_T$
if no polynomial-time algorithm has advantage at least $\epsilon$ in solving the decision $k$-BDHE problem in $\mathbb{G}_T$.

	\section{Non-interactive Key Exchange Protocol}
	
	In this section, we first give a high-level description of our dynamic group non-interactive key exchange protocol and then present the instantiated protocol.
	\subsection{High-level Description}
	Our NI-CBE protocol is defined by the following algorithms:
	\begin{itemize}
		\item {\sf GlobeSetup}: This algorithm is used to generate the global system parameter.
		\item {\sf KeyRegis}: This algorithm allows each user in the open network environment to register with a trusted authority (TA) and get a long-term public-private key pair.
		\item {\sf KeyDerive}: This algorithm allows multiple users to form a group and negotiate a shared group encryption key and their respective decryption keys in a non-interactive way.
		\item {\sf KeyUpdate}: This algorithm is used to update the group encryption key and each group member's decryption key once the group membership changes.
		\item {\sf Encrypt}: This algorithm allows any sender knowing the group encryption key of a group to send an encrypted session key (also called the broadcast ciphertext) to some chosen members within the group (often called recipients).
		\item {\sf Decrypt}: This algorithm allows a recipient to obtain the session key through decrypting the broadcast ciphertext using its decryption key.
	\end{itemize}
	
	\subsection{The Dynamic Group Non-interactive Key Exchange Protocol} \label{Scheme}
	Our concrete NI-CBE protocol comes as follows:

	\begin{itemize}
		\item {\sf GlobeSetup}($1^\lambda$): On input the security parameter $1^\lambda$, it generates a public system parameter list ${\sf params}$ as follows: choose two cyclic multiplicative groups  $\mathbb{G}_1,\mathbb{G}_T$ with prime order $q$, where $\mathbb{G}_1$ is generated by $g$; choose a bilinear map $\hat{e}: \mathbb{G}_1 \times \mathbb{G}_1  \rightarrow \mathbb{G}_T$, $u \in \mathbb{G}_1$ and generate $\mathbb{H}=\{h_1,h_2,...,h_l\}$ by randomly selecting $h_i\in \mathbb{G}_1, 1\leq i \leq l$; generate $L_s$ tuples of the format $(\mathbb{A}_\gamma,\mathbb{B}_\gamma,\mathbb{K}_\gamma)$, $1\leq \gamma \leq L_s$. For each tuple, it corresponds to a group with the maximal group size $n$ and is generated as follows:
		\begin{itemize}
			\item For $1\leq i\leq n$, choose $\alpha_{i\gamma},\beta_{i\gamma} \in \mathbb{Z}_q^*$, compute $A_{i\gamma}=g^{\alpha_{i\gamma}}$, $B_{i\gamma}=g^{\beta_{i\gamma}}$ and set $A_\gamma=\{A_{i\gamma}\}_{i\in \{1,...,n\}}$ and $B_\gamma=\{B_{i\gamma}\}_{i\in \{1,...,n\}}$.
			\item For $1\leq i,j\leq n, i \neq j$, compute $K_{ij\gamma}=h_j^{\alpha_{i\gamma}}u^{\beta_{i\gamma}}$ and set $K_{j\gamma}=\{K_{ij\gamma}\}_{1\leq i \leq n, i \neq j}$, $K_\gamma=\{K_{j\gamma}\}_{1\leq j \leq n}$.
		\end{itemize}
		Finally, ${\sf params}=(\mathbb{G}_1,\mathbb{G}_T, \mathbb{Z}_q^*, \hat{e}, g,q, u, \mathbb{H}, \{\mathbb{A}_\gamma, \mathbb{B}_\gamma,\mathbb{K}_\gamma\}_{1\leq \gamma \leq L_s})$.
		Assume there are totally $N$ users in the open networks. Let $\mathbb{P}=\{P_1,P_2,...,P_N\}$ denote all the users. Then we have a set $\mathbb{I}=\{1,2,...,N\}$ recording the index of each user in $\mathbb{P}$. This index is called user index. 
		
		\item {\sf KeyRegis}: Suppose that a party $P$ with user index $i'$ intends to obtain a long-term public-private key pair from TA. 
		
		\smallskip
		{\sf Case 1}: {\sf KeyRegis}($(\mathbb{A}_\gamma,\mathbb{B}_\gamma,\mathbb{K}_\gamma),i$): For a user, the index within a group is called group index. In this case, $P$ determines its group index as $i$ and the group corresponds to a tuple $(\mathbb{A}_\gamma,\mathbb{B}_\gamma,\mathbb{K}_\gamma)$. Then, TA generates a long-term public-private key pair ($PK_i,SK_i$) for $P$: choose $a_i,b_i \in \mathbb{Z}_q^*$ randomly; compute $A_i=g^{a_i}, B_i=g^{b_i}, K_{ij}=h_j^{a_i}u^{b_i}$, $1\leq j \leq n$; get $SK_i=h_i^{a_i}u^{b_i}$, $PK_i=(i,A_i,B_i, \{K_{ij}\}_{1\leq j \leq n, j\neq i})$.

		\smallskip
		{\sf Case 2}: {\sf KeyGen}($\mathbb{A}_\gamma,\mathbb{B}_\gamma,\mathbb{K}_\gamma$): In this case, $P$ doesn't know its group index, TA firstly generates $n$ pairs of public-private keys $\{PK_l,SK_l\}$ as follows: for $1\leq l \leq n$, select $a_l,b_l \in \mathbb{Z}_q^*$, get $A_l=g^{a_l}, B_l=g^{b_l}, K_{lj}=h_j^{a_l}u^{b_l}$, $1\leq j \leq n$; for $1\leq l \leq n$, set $SK_l=h_l^{a_l}u^{b_l}$, $PK_l=(l,A_l,B_l, \{K_{lj}\}_{1\leq j \leq n, j\neq l})$. 
		Assume $P$ is the first user to enroll with TA, TA assigns the first public-private key pair to $P$, which implies $P$ occupies the first position in a group corresponding to ($\mathbb{A}_\gamma,\mathbb{B}_\gamma,\mathbb{K}_\gamma$).

		For simplicity, we regulate {\sf KeyRegis} in our protocol runs as in {\sf Case 1}. We note in both cases, TA issues a certificate to each legitimate user so as to ensure the validity of the user's public key.


		\item {\sf KeyDerive}($(\mathbb{A}_\gamma,\mathbb{B}_\gamma,\mathbb{K}_\gamma),i,SK_i,\mathbb{U},\{PK_i\}_{i\in\mathbb{U}},id_\pi$): Assume there are $t$ users intending to establish a group corresponding to $(\mathbb{A}_\gamma,\mathbb{B}_\gamma,\mathbb{K}_\gamma)$. Let $\pi$ and $id_\pi$ denote the index of the group and a unique identifier of this group. We note $id_\pi$ can be randomly chosen by one of the users. Assume these users' indexes within the group form an index set $\mathbb{U}=\{1,...,t\}$.  For $i\in \mathbb{U}$, each user $P$ with the index within the group $i$ and its private-public key pair $(SK_i,PK_i)$ performs the following steps to obtain a shared group encryption key $\Omega$ and a decryption key $d_i$:
		\begin{itemize}
			\item For $i\in \mathbb{U}$, parse $PK_i$ as $(i,A_i,B_i, \{K_{ij}\}_{1\leq j \leq n, j\neq i})$.
			\item Set $Y_1=\prod_{i=1}^{t}A_i\prod_{i=t+1}^{n}A_{i\gamma}$, $Y_2=\prod_{i=1}^{t}B_i\prod_{i=t+1}^{n}B_{i\gamma}$ and output the group encryption key $\Omega=(Y_1,Y_2)$.
			\item For $1\leq i \leq n$, set  $\hat{dk}_i=\prod_{j=1}^{t,j\neq i}K_{ji}\prod_{j=t+1}^{n,j\neq i}K_{ji\theta}$.
			\item Set $d_i=\hat{dk}_iSK_i$. If $\hat{e}(d_i,g)\stackrel{?}{=}\hat{e}(h_i,Y_1)\hat{e}(u,Y_2))$, output $d_i$ as the decryption key; else, abort the algorithm.
		\end{itemize}
		Apart from outputting ($\Omega, d_i$), the algorithm also outputs $(G_\pi,M_{\pi}^{i})$. $G_\pi=(\pi,id_\pi, \Omega, st,\Delta)$ is used to describe some basic information about a group, which can be accessed publicly. We note $st$ is an $n$-bit string initialized with all zero, which is used to record all the positions occupied by all the group members. That is, if the $i$-th position is occupied by a group member, set $[st]_i=1$. $\Delta$ denotes an index set recording all the group indexes of group members. 
		$M_{\pi}^i=(\hat{dk}_1,...,\hat{dk}_n,d_i)$ denotes the $i$-th group member's member information corresponding to the $\pi$-th group, which is stored in the member's local database.

		\item {\sf KeyUpdate}($G_\pi,PK$): The algorithm allows a user/member with $PK$ to join/\\leave a group with $G_\pi$ in a non-interactive manner. The algorithm can be discussed in the following two cases:

		\smallskip
		{\sf Join}: Assume a user $P_{i'}$ wants to join a group with $G_\pi$ as the $I$-th group member, and currently there are $t$ group members that form an index set $\{1,...,t\}$. Let $PK_I$ denote the public key of $P_{i'}$. If $([st]_I=1)\vee (t+1> n)$, the algorithm aborts; else, it performs as follow:
		\begin{itemize}
			\item Parse $PK_I$ as $(I,A_I,B_I, \{K_{Ij}\}_{1\leq j \leq n, j\neq I})$;
			\item Compute the group encryption key $\Omega$ and a decryption key $d_I$ for the new $I$-th group member by invoking {\sf KeyDerive} with inputs $((\mathbb{A}_\gamma,\mathbb{B}_\gamma,\mathbb{K}_\gamma),I,\\SK_i,\mathbb{V},\{PK_j\}_{j\in \mathbb{V}}, id_\pi)$, where $\mathbb{V}=\{1,...,t+1\}$ and $I\in \mathbb{V}$.
			\item Update the old group encryption key $\Omega$ and the string $st$ in $G_\pi$ by setting $Y_1=Y_1A_IA_{I\gamma}^{-1}$, $Y_2=Y_2B_IB_{I\gamma}^{-1}$ and $[st]_I=1$;
			\item For $1\leq j\leq t, j\neq I$, update each old member's member information $M_{\pi}^j$ by setting $\hat{dk}_l=\hat{dk}_lK_{Il}K_{Il\gamma}^{-1}$, $1\leq l\neq I \leq n$, and  $d_j=d_jK_{Ij}K_{Ij\gamma}^{-1}$;
			\item Add $i'$ to $\Delta$ and output the new group information $G_\pi$.
			
		\end{itemize}
		
		We note for the $I$-th new member, it accepts $d_I$ as its decryption key iff the equation is satisfied: $\hat{e}(d_I,g)\stackrel{?}{=}\hat{e}(h_I,Y_1)\hat{e}(u,Y_2))$.
		
		\smallskip
		{\sf Leave}: Assume there are currently $t$ group members who form an index set $\{1,...,t\}$ in the $\pi$-th group, and the $J$-th group member with user index $i'$ and public key $PK_J$ wants to leave this group permanently. The algorithm performs as follows:
		\begin{itemize}
			\item 	Parse $PK_J$ as $(J,A_J,B_J, \{K_{Jj}\}_{1\leq j \leq n, j\neq J})$;
			\item Update the group encryption key $\Omega$ and the string $st$ in $G_\pi$ by setting $Y_1=Y_1A_J^{-1}A_{J\gamma}$, $Y_2=Y_2B_J^{-1}B_{J\gamma}$ and $[st]_J=0$;
			\item For $1\leq j \leq (t-1), j\neq J$, update the remaining member's member information $M_\pi^j$ by setting $\hat{dk}_l=\hat{dk}_lK_{Jl\gamma}K_{Jl}^{-1}$, $1\leq l, J\leq n, l\neq J$ and $d_j=d_jK_{Jj}^{-1}K_{Jj\gamma}$.
			\item Remove $i'$ from $\Delta$ and output the new group information $G_\pi$.
		\end{itemize}
		
		For the rest of member, it accepts its new decryption key $d_j$ iff  $\hat{e}(d_j,g)\stackrel{?}{=}\hat{e}(h_j,Y_1)\hat{e}(u,Y_2))$.

		\item {\sf Encrypt($G_\pi, \Delta',\mathbb{U}$)}: Anyone knowing the public group information could run the algorithm. Assume a user $P$ chooses a group with $G_\pi$ and some group members within the group whose user index set is denoted as $\Delta'\subseteq \Delta$. Let $\mathbb{U}$ and $\mathbb{S}=\{i|\forall i \in \{1,...,n\}, [st]_i=1\}$ respectively denote the group index set of chosen group members and the group index set of all the group members in the $\pi$-th group, where $\mathbb{U}\subseteq \mathbb{S}$. On input $(G_\pi,\mathbb{U})$, the algorithm performs as follows:
		
		
		\begin{itemize}
			\item Get a set $\bar{\mathbb{U}}=\mathbb{S}\setminus \mathbb{U}$ and compute $\hat{Y}_1=Y_1\prod_{i\in \bar{\mathbb{U}}}A_{i\gamma}$, $\hat{Y}_2=Y_2\prod_{i\in \bar{\mathbb{U}}}B_{i\gamma}$;
			
			\item Choose $\rho$ from $ \mathbb{Z}_q^*$ at random and compute  $C_1=g^\rho$, $C_2=\hat{Y_1}^\rho$;

			\item Get a set $\bar{\mathbb{S}}=\{i|\forall i\in \{1,...,n\}, [st]_i=0\}$ and compute a session key $k=\hat{e}(u^\rho, \prod_{i\in \mathbb{S}}B_i \prod_{i\in \mathbb{\bar{U}}}B_{i\gamma}\prod_{i\in \mathbb{\bar{S}}}B_{i\gamma})$;
			
			\item Output a pair $(C_h,k)$, where $C_h=(C_1,C_2)$.
		\end{itemize}
		
		We note the {\sf Encrypt} algorithm finally outputs a pair ($C_h,k$), where $C_h$ is called the header and $k\in \mathbb{G}_T$ is essentially a message encryption key. In the sequel, once $P$ shares a session key $k$ with group members in $\mathbb{U}$, $P$ will choose a message $m$
		from the message space and encrypt $m$ under $k$ by using a semantically-secure symmetric encryption scheme. In addition, for the convenience of decryption for chosen group members, a sender generally uploads the tuple $(P,id_\pi,\Delta',\mathbb{U},C_h)$ to a platform(i.e., a server). 
		
		\item {\sf Decrypt($G_\pi,\Delta',i',\mathbb{U},d_i,C_h$)}: For any user $P_{i'}$ with user index $i'$, if $i'\in\Delta'$, then $P_{i'}$ computes the same session key with the sender by invoking the algorithm. On input ($G_\pi,\Delta,\mathbb{U},i',d_i,C_h$), if $i\notin \mathbb{U}$, the algorithm outputs $null$; otherwise, it outputs the session key as follows:
		\begin{itemize}
			\item Parse $C_h$ as $(C_1,C_2)$;
			\item Compute $\hat{d}_i=d_i\prod_{l\in \mathbb{\hat{U}}}K_{li\gamma}$;
			\item Compute and output the session key $k=\hat{e}(\hat{d_i},C_1)\hat{e}(h_i, C_2)^{-1}$.
		\end{itemize}
		
		We then show our NI-CBE protocol satisfies correctness. That is, if a sender gets a pair ($C_h,k$) by invoking {\sf Encrypt} with $(G_\pi,\Delta',\mathbb{U})$ as inputs. For each group member ($i\in \mathbb{U}$), it can compute the same session key $k$ with the sender by invoking {\sf Decrypt} algorithm with inputs $(G_\pi,\Delta',\mathbb{U},i,d_i,C_h)$. We note the correctness of NI-CBE protocol is guaranteed by the correctness of the equation in {\sf Decrypt} which is shown below:
		\begin{align*}
			\hat{e}(\hat{d}_i,C_1)\hat{e}(h_i,C_2)^{-1}
			&=
			\hat{e}(\prod_{l\in\mathbb{S}}K_{l},g^\rho)\hat{e}(\prod_{l\in\mathbb{\bar{S}}}K_{li\gamma},g^\rho) \hat{e}(\prod_{l\in\mathbb{\bar{U}}}K_{li\gamma},g^\rho) \hat{e}(h_i^\rho,Y_1\prod_{l\in \mathbb{\bar{U}}}A_{l\gamma})^{-1}
			\\ &= \hat{e}(\prod_{l\in\mathbb{S}}h_i^{a_l}u^{b_l},g^\rho)
			\hat{e}(\prod_{l\in\mathbb{\bar{S}}}h_i^{\alpha_{l\gamma}}u^{\beta_{l\gamma}},g^\rho) \hat{e}(\prod_{l\in\mathbb{\bar{U}}}h_i^{\alpha_{l\gamma}}u^{\beta_{l\gamma}},g^\rho) \\ & \hat{e}(h_i^\rho,\prod_{l\in \mathbb{S}}A_{l}\prod_{l\in \mathbb{\bar{S}}}A_{l\gamma}\prod_{l\in \mathbb{\bar{U}}}A_{l\gamma})^{-1}
			\\ & = \hat{e}(h_i^\rho,\prod_{l\in\mathbb{S}}A_l \prod_{l\in\mathbb{\bar{S}}}A_{l\gamma} \prod_{l\in\mathbb{\bar{U}}}A_{l\gamma}) \hat{e}(u^\rho,\prod_{l\in \mathbb{S}}g^{b_l}\prod_{l\in \mathbb{\bar{S}}}g^{\beta_{l\gamma}} \prod_{l\in \mathbb{\bar{U}}}g^{\beta_{l\gamma}})
			\\  & \hat{e}(h_i^\rho,\prod_{l\in \mathbb{S}}R_{l}\prod_{l\in \mathbb{\bar{S}}}A_{l\gamma}\prod_{l\in \mathbb{\bar{U}}}A_{l\gamma})^{-1}
			\\ & =\hat{e}(u^\rho,\prod_{l\in \mathbb{S}}g^{b_l}\prod_{l\in \mathbb{\bar{S}}}g^{\beta_{l\gamma}} \prod_{l\in \mathbb{\bar{U}}}g^{\beta_{l\gamma}})
			\\ & = \hat{e}(u^\rho,\prod_{l\in \mathbb{S}}B_l\prod_{l\in \mathbb{\bar{S}}}B_{l\gamma} \prod_{l\in \mathbb{\bar{U}}}B_{l\gamma})
		\end{align*}

	\end{itemize}

	\section{Security Analysis of NI-CBE Protocol}
	In this section, we first design the security model for our NI-CBE protocol and then we give the formal security proof of our protocol.
	\subsection{Security Model}
	We have defined a user set $\mathbb{P}=\{P_1,P_2,...,P_N\}$ and an user index set $\mathbb{I}=\{1,2,...,N\}$. It is known that any users who form a user set $\mathbb{\hat{U}}$ ($\mathbb{\hat{U}} \subseteq \mathbb{P}$) can establish a group in the non-interactive way. Each built group has an identity $id_\pi$ and a session round $\ell$, where $\pi$ denotes it is the $\pi$-th group currently. For a group which is firstly formed by a group of users, the session round $\ell$ is set to be $1$. If a user/member joins/leaves the group, the session round $\ell$ will increase by $1$. In our security model, the group information describing each formed group is denoted as 
	$G_{\pi}=(\pi,\ell,id_\pi,st,\Omega,\Delta)$, which corresponds to the $\ell$-th session round of the $\pi$-th group. In the sequel, we give the security model of our G-NIKE. It is essentially a security game run between a challenger $\mathcal{C}$ and a probabilistic polynomial time (PPT) adversary $\mathcal{A}$. The security game  consists of three phases as follows:

	\noindent \textbf{Initial}: In this phase, $\mathcal{C}$ generates $\sf params$ by running {\sf GlobeSetup} algorithm with a security parameter $\lambda$,  and returns $\sf params$ to $\mathcal{A}$. $\mathcal{A}$ then submits a subset of  $\mathbb{I}$ to $\mathcal{C}$, which is denoted as $\mathbb{U}$. This phase is used to simulate {\sf GlobeSetup} algorithm.

	\noindent \textbf{Query}:
	In the second phase, $\mathcal{C}$ answers the following types of queries from $\mathcal{A}$:
	\begin{itemize}
		\item Register($(\mathbb{A}_\gamma,\mathbb{B}_\gamma,\mathbb{K}_\gamma),i',i$): This query is used to model {\sf KeyRegis} algorithm, which prompts $\mathcal{C}$ to register a user selected by $\mathcal{A}$. In particular, $\mathcal{C}$ maintains an initially empty list $\mathbf{L}_u$. $\mathcal{A}$ chooses and supplies a user index $i'$ from $\mathbb{I}$. Assume $\mathcal{A}$ submits the $\gamma$-th tuple corresponding the group and the group index $i$ of this user. $\mathcal{C}$ first generates a public-private key pair $(PK_i,SK_i)$, then adds $(i',PK_i,SK_i)$ to $\mathbf{L}_u$. Finally, $\mathcal{C}$ replies $PK_i$ to $\mathcal{A}$.
		

		\item Extract($i'$): This query allows $\mathcal{A}$ to extract the long-term private key held by a user who has been registered. $\mathcal{A}$ first inputs an index $i'$ of a user $P_{i'}$. If $i'\in \mathbb{U}$, $\mathcal{C}$ aborts; else, $\mathcal{C}$ recovers the item $(i',PK_i,SK_i)$ from $\mathbf{L}_u$ and returns $SK_i$ to $\mathcal{A}$.
		
		\item Execute($(\mathbb{A}_\gamma,\mathbb{B}_\gamma,\mathbb{K}_\gamma),i,SK_i,\hat{\mathbb{S}},\{PK_j\}_{j\in \hat{\mathbb{S}}},id_\pi,\ell$): This query is used to model {\sf KeyDerive} algorithm. $\mathcal{A}$ asks $\mathcal{C}$ to form a group for multiple users with a user index set $\mathbb{S}\subseteq \mathbb{I}$ and obtains the group encryption key and one of these users' decryption key. Assume the user $\mathcal{A}$ designates has the group index $i$, the group index set for these users is $\hat{\mathbb{S}}$, and it is the $\ell$-th session of the $\pi$-th group. If $\mathbb{\hat{S}}\wedge \mathbb{U}\ne \emptyset$, $\mathcal{C}$ aborts; else, $\mathcal{C}$ generates and replies the group encryption key $\Omega$ and the decryption key $d_i$ held by the $i$-th group member in the group.
		
		\item Join($i',PK_i,G_\pi,I$):  This query is used to model {\sf Join} sub-algorithm of {\sf KeyUpdate} algorithm, which allows $\mathcal{A}$ to randomly choose a user who has been registered previously to join a group. Assume $\mathcal{A}$ selects a group with $G_\pi$ that corresponds to an initial tuple $(\mathbb{A}_\gamma,\mathbb{B}_\gamma,\mathbb{K}_\gamma)$ and the $I$-th position in the $\pi$-th group to add a new group member with a user index $i'$ and a corresponding public key $PK_I$. If the query is invoked successfully, $\mathcal{C}$ replies the updated $G_\pi$ to $\mathcal{A}$.
		
		\item  Leave($PK_I,G_\pi,I$): This query is used to model {\sf Leave} sub-algorithm of {\sf KeyUpdate} algorithm, which allows $\mathcal{A}$ to choose any group member to leave a group permanently. Assume $\mathcal{A}$ selects a group with $G_\pi$ and the $I$-th group member within the group, and  the $I$-group member has a public key $PK_I$.  If the query is invoked successfully, $\mathcal{C}$ replies the updated $G_\pi$ to $\mathcal{A}$.
		
		\item Reveal($i',\pi,\ell$): This query allows $\mathcal{A}$ to obtain the decryption key held by any user who participates in the $\ell$-th session of the $\pi$-th group. Assume the user has user index $i'$ and group index $i$. If $i'\in \mathbb{U}$, $\mathcal{C}$ aborts; else, $\mathcal{C}$ returns the decryption key $d_i$ held by $i$-th group member in the group. 
		
		\item Test($\mathbb{U^*},id_{\pi^*},\ell^*$): In this query, $\mathcal{A}$ first chooses a target group with $id_{\pi^*}$, a target session round $\ell^*$ of the $\pi^*$-th group and a target user index set $\mathbb{U}^*$ ($\mathbb{U}^*\subseteq \mathbb{U}$) within the target group. Suppose the group information of the target group is denoted as $G_{\pi^*}=\{\pi^*,id_{\pi^*},\ell^*,st,\Omega,\Delta\}$.
		On input ($\mathbb{U^*},id_{\pi^*},\ell^*$), $\mathcal{C}$ tosses a coin $b\in \{0,1\}$ firstly. If $b=0$, $\mathcal{C}$ gets $C_{h^*}$ and a real session key $k_0$ by invoking {\sf Encrypt} algorithm; else, $\mathcal{C}$ selects a random session key $k_1$ from the session key space $\mathbb{G}_T$. At last, $\mathcal{C}$ replies $(C_{h^*},k_b)$ to $\mathcal{A}$.
	\end{itemize}
	
	\noindent \textbf{Guess}:
	In this phase, $\mathcal{A}$ submits $b'\in \{0,1\}$ to $\mathcal{C}$. If $\mathcal{A}$ is able to distinguish a valid session key calculated by a set of users from a random element of the session key space, that is $b'=b$, then $\mathcal{A}$ wins the above game with advantage ${Adv}_{\mathcal{A}}$, where $Adv_{\mathcal{A}}=2|Pr[b'=b]-1|$.
	
	
	\begin{definition} \label{Semi-secure} Our dynamic group non-interactive key exchange protocol is semi-statically secure if for any PPT adversary $\mathcal{A}$ in the above game satisfying the following conditions, the advantage $Adv_{\mathcal{A}}$ of $\mathcal{A}$ to win the above game is negligible.
		
		\begin{itemize}
			
			\item $\mathcal{A}$ submits a user index set $\mathbb{U}$ after obtaining the public system parameter.
			\item Each query to Extract must be on an index $i'$ outside $\mathbb{U}$.
			\item Each query to Reveal must have $\mathbb{\hat{U}}\wedge \mathbb{U}= \emptyset$.
			\item The query to Test must be on a subset $\mathbb{U}^*$ of $\mathbb{U}$.
		\end{itemize}
	\end{definition}

	\subsection{Security Proof}
	In this section, we propose the following theorem and the corresponding proof to present that our NI-CBE protocol is semi-statically secure in above security game.
	\begin{theorem}\label{thm}
		Assume that there are at most $N$ groups which can be established by invoking our NI-CBE protocol, and for each group, there are at most $L$ sessions that can be launched. If there exists an adversary $\mathcal {A}$ who wins the above security game with advantage $\epsilon$, then there exists an algorithm
		to solve the decision $k$-BDHE problem with advantage $\frac{1}{NL}\epsilon$.
	\end{theorem}

	\noindent\emph{Proof}. Suppose $\mathcal {C}$ is given an instance
	$(g,Q,Z,X_1,...,X_k,X_{k+2},...,X_{2k})$ of the decision $k$-BDHE problem,
	where $X_i=g^{\theta^i}, i\in \{1,...,k,k+2,...,2k\}$ with an
	unknown $\theta\in \mathbb{Z}_q$. We show how $\mathcal {C}$ can use $\mathcal {A}$ to determine whether
	$Z$ equals to $\hat{e}(g^{\theta^{k+1}},Q)$ or a uniform element in $\mathbb{G}_T$.

	\noindent \textbf{Initial}: $\mathcal {C}$ generates ${\sf params}$ as follows: 
	for $1\leq j \leq n$, choose $\zeta_j \in \mathbb{Z}_q^*$ and set $h_j=g^{\zeta_j}X_j $; set $P=g^{\alpha^k}=X_k$; for $1\leq \gamma \leq L_s$, generate  $(\mathbb{A}_\gamma, \mathbb{B}_\gamma, \mathbb{K}_\gamma)$ that corresponds to the maximal group size $n$:

	\begin{itemize}
		\item If $i= 1$, select $\alpha_{1\gamma},\beta_{1\gamma}\in \mathbb{Z}_q^*$ randomly and compute
		$A_{1\gamma}=g^{\alpha_{1\gamma}}\prod_{i=2}^{n} X_{k-i+1}^{-1}, $
		$B_{1\gamma}=g^{\beta_{1\gamma}}y_1,$
		set $K_{1j{\gamma}}=A_{1\gamma}^{\zeta_j}X_j^{\alpha_{1\gamma}}\prod_{i=2}^{n}X_{k-l+1+j}^{-1}P^{\beta_{1\gamma}}$ for $2\leq j\leq n$,
		set $K_{11\gamma}=\perp$.
		
		\item Else ($2\leq i\leq n$), select $\alpha_{i\gamma},\beta_{i\gamma}\in \mathbb{Z}_q^*$ randomly, compute $A_{i\gamma}=g^{\alpha_{i\gamma}} X_{k-i+1}, $ $B_{i\gamma}=g^{\beta_{i\gamma}}$, set $K_{ij\gamma}=A_{1\gamma}^{\zeta_j }X_j^{\alpha_{i\gamma}}X_{k-i+1+j} P^{\beta_{i\gamma}}$ for $2\leq j \leq n$, set $K_{ii\gamma}=\perp$.

	\end{itemize}
	$\mathcal{C}$ returns ${\sf params}=(\mathbb{G}_1,\mathbb{G}_T, \mathbb{Z}_q^*, \hat{e}, g,q, P, \mathbb{H}, \{\mathbb{A}_\gamma, \mathbb{B}_\gamma,\mathbb{K}_\gamma\}_{1\leq \gamma \leq L_s})$ to $\mathcal{A}$.
	$\mathcal{A}$ then submits a user index set $\mathbb{U}\subseteq \mathbb{I}$ to $\mathcal {C}$. 
	
	Assume there are at most $N$ groups that have been formed by $\mathcal {C}$ and in each group, the maximal number of session rounds is $L$. 
	We note $\mathcal{C}$ chooses a group from all $N$ groups as a target group (assume the $\pi$-group)and a corresponding target session round $\ell$ in advance. We note if it's not the $\pi$-th group, then all the transcripts are consistent with that in the real protocol, which means $\mathcal {C}$ can answer all the following queries correctly. In other words, we only need to consider the queries associated with the target group.

	\medskip
	\noindent \textbf{Query}: $\mathcal{C}$ answers the following queries from $\mathcal{A}$:

	\noindent {\sf Register($(\mathbb{A}_\gamma,\mathbb{B}_\gamma,\mathbb{K}_\gamma),i',i$)}: $\mathcal{C}$ maintains an initially empty list $\mathbf{L}_u$. To answer the query, $\mathcal{C}$ performs as follows:
	
	
	\begin{itemize}
		\item If there exists an item $(i',PK_i,SK_i)$ on $\mathbf{L}_u$, return $PK_i$ as the answer;
		\item Else, select $a_i,b_i$ from $\mathbb{Z}_q^*$
		and do the following:
		\begin{itemize}
			\item If $i' \notin \mathbb{U}$, set $A_i=g^{a_i}$,$B_i=g^{b_i}$, $K_{ij}=h_j^{a_i}P^{b_i}$, for $1\leq j\neq i \leq n$; set $PK_i=(i,A_i,B_i,\{K_{ij}\}_{1\leq j \leq n,j\neq i})$, $SK_i=K_{ii}$; add the item $(i',PK_i,SK_i)$ to $\mathbf{L}_u$ and return $PK_i$ as the answer.
			\item Else, if $i\neq n$, set $A_{i}=g^{a_{i}} X_{k-i+1}$, $B_{i}=g^{b_{i}}$, $K_{ij}=A_i^{\zeta_j}X_j^{a_{i}}X_{k-i+1+j}P^{b_{i}} $, $1\leq j\neq i \leq n$, set $PK_i=(i,A_i,B_i,\{K_{ij}\}_{1\leq j \leq n,j\neq i} )$, $SK_i=\perp$; else, set 
			$A_{i}=g^{a_i}\prod_{i=2}^{n} X_{k-i+1}^{-1}$,
			$B_{i}=g^{b_i}X_1$,
			$K_{ij}=A_i^{\zeta_j}X_j^{a_i}
			\prod_{i=2}^{n}x_{k-i+1+j}^{-1}P^{d_i}$, $1\leq j\neq i \leq n$, set $PK_i=(i,A_i,B_i,\{K_{ij}\}_{1\leq j \leq n,j\neq i} )$, $SK_i=\perp$; add the item $(i',PK_i,SK_i)$ to $\mathbf{L}_u$ and return $PK_i$ as the answer.
		\end{itemize}
	\end{itemize}

	\noindent {\sf Extract($i'$)}: On receiving a user index $i'$, $\mathcal{C}$ does the following: if $i' \in \mathbb{U}$, abort; else, recover the item $(i',PK_i,SK_i)$ from $\mathbf{L}_u$ and return $SK_i$ to $\mathcal{A}$.

	\noindent {\sf Execute($(\mathbb{A}_\gamma, \mathbb{B}_\gamma,\mathbb{K}_\gamma),i,SK_i,\hat{\mathbb{S}},\{PK_j\}_{j\in \hat{\mathbb{S}}},id_\pi,\ell$)}: Assume currently it is the $\ell$-th session of $\pi$-th group and the user index set is $\Delta$ that corresponds to a group index set $\hat{\mathbb{S}}=\{1,...,t\}$. If $\mathbb{\hat{S}}\wedge \mathbb{U}\ne \emptyset$, $\mathcal{C}$ aborts; else, $\mathcal{C}$ does the following:
	\begin{itemize}
		\item Compute $Y_1=\prod_{l=1}^{t}A_l\prod_{l=t+1}^{n}A_{l\gamma}$, $Y_2=\prod_{l=1}^{t}B_l\prod_{l=t+1}^{n}B_{l\gamma}$ and get the group encryption key $\Omega=(Y_1,Y_2)$ of $\pi$-th group.
		\item Recover $SK_i$ from $\mathbf{L}_u$, for $1\leq i \leq n$, compute $\hat{dk_i}=\prod_{l=1}^{t,l\neq i}K_{li} \prod_{l=t+1}^{n,l\neq i}K_{li\gamma}$ and get the decryption key held by $i$-th group member $d_i=\hat{dk}_iSK_i$. 
		\item Generate $n$-bit empty string $st$. For $1\leq i \leq n$, if $i\in \mathbb{\hat{S}}$, set $st_i=1$. 
		\item Generate the member information for each group member: for $1\leq i \leq t$, get $M_{\pi,\ell}^i=\{i',i,\hat{dk}_1,...,\hat{dk}_n,d_i\}$.
		\item Return the group information of $\pi$-th group $G_\pi=(\pi,\ell,id_\pi,st,\Omega,\Delta)$.
	\end{itemize}

	After answering the query, $\mathcal{C}$ generates a list $\mathbf{T}_{\pi,\ell}=(G_\pi,\{M_{\pi,\ell}^{i}\}_{1\leq i \leq t})$, which corresponds to the $\ell$-th session of the $\pi$-th group. We note by invoking the following {\sf Join} or {\sf Leave} query, the number of group members will increase or decrease correspondingly. If $\mathcal{C}$ sets $\mathbf{T}_{\pi,\ell}=\mathbf{T}_{\pi,\ell-1}$ in the sequel {\sf Join} or {\sf Leave} query, $\mathcal{C}$ does the following: 1) Replace $G_\pi$ with the updated $G_\pi$; 2) For $i\in \{1,..,t\}$, set $M_{\pi,\ell}^{i}=M_{\pi,\ell-1}^{i}$.

	\noindent {\sf Join($i',PK_i,G_\pi,I$)}: Assume the current session round is $\ell$, there exists $t$ group members currently, the current group information is $G_\pi=\{\pi,id_\pi,\ell,st,\Omega,\Delta\}$, and the group index of all group members of the $\pi$-th group forms a group index set $\mathbb{V}=\{1,...,t\}$. If $st[I]\neq \perp$ or $t+1 > n$, $\mathcal{C}$ aborts; else, $\mathcal{C}$ first recovers the tuple $(i',PK_i,SK_i)$ from $\mathbf{L}_u$ and then does the following:
	\begin{itemize}
		\item Set $\mathbf{T}_{\pi,\ell}=\mathbf{T}_{\pi,\ell-1}$ and parse $PK_i$ as $(i,A_i,B_i,\{K_{ij}\}_{1\leq j \leq n,j\neq i} )$.
		\item Update $\Omega$ by setting $Y_1=Y_1A_{i\gamma}^{-1}A_{i}$ and  $Y_2=Y_2B_{i\gamma}^{-1}B_{i}$. 
		\item For $j \in \mathbb{V}$, update $M^j_{\pi,\ell}$ by setting $\hat{dk}_l=\hat{dk}_l K_{jl\gamma}^{-1}K_{jl}$, for $1\leq l \neq j \leq n$ and $d_j=d_jK_{ji\gamma}^{-1}K_{ji}$. 
		\item Set $st[I]=1$, 
		and $M^{I}_{\pi,\ell}=\{I,\hat{dk}_1,...,\hat{dk}_n,d_I\}$ and add $M^{I}_{\pi,\ell}$ to the list $\mathbf{T}_{\pi,\ell}$, where $d_I=\hat{dk}_iSK_i$. 
		\item Add $i'$ to $\Delta$ and return the updated $G_\pi$ to $\mathcal{A}$. 
	\end{itemize}

	\noindent {\sf Leave($PK_I,G_\pi,I$)}: Assume the current session round is $\ell$, there are $t$ existing group members totally, and the group information is $G_\pi=\{\pi,id_\pi,\ell,st,\Omega,\Delta\}$. Assume the group index of all group members of the $\pi$-th group forms a group index set $\mathbb{V}=\{1,...,t\}$. 
	To update $G_\pi$, $\mathcal{C}$ performs as follows:
	\begin{itemize}
		\item Set $\mathbf{T}_{\pi,\ell}=\mathbf{T}_{\pi,\ell-1}$ and parse $PK_I$ as $(I,A_I,B_I,\{K_{Ij}\}_{1\leq j \leq n,j\neq I} )$. 
		\item Update $\Omega$ by setting $Y_1=Y_1A_{I\gamma}A_{I}^{-1}$ and $Y_2=Y_2B_{I\gamma}B_{I}^{-1}$. 
		\item For $i \in \mathbb{V}$, update $M^i_{\pi,\ell}$ by setting $\hat{dk}_l=\hat{dk}_l K_{Il\gamma}K_{Il}^{-1}$, for $1\leq l \neq I \leq n$ and $d_i=d_iK_{Ii\gamma}K_{Ii}^{-1}$. 
		\item Set $st[I]=\perp$ and remove $M^I_{\pi,\ell}$ from the list $\mathbf{T}_{\pi,\ell}$. 
		\item Removes $i'$ from $\Delta$ and return the updated $G_\pi$ to $\mathcal{A}$. 
	\end{itemize}

	\noindent {\sf Reveal($i',\pi,\ell$)}: If $i'\notin \mathbb{U}$, $\mathcal{C}$ recovers $M_{\pi,\ell}^i$ from $\mathbf{T}_{\pi,\ell}$ and returns $d_i$ to $\mathcal{A}$; Else, $\mathcal{C}$ this query. 
	
	
	

	\noindent {\sf Test($\mathbb{U}^*,id_{\pi^*},\ell^*$)}:  $\mathcal{A}$ submits a target group with $id_{\pi^*}$, a target session round $\ell^*$ and a user index set $\mathbb{U}^*\subseteq \mathbb{U}$. Suppose $(\mathbb{A}_\gamma, \mathbb{B}_\gamma,\mathbb{K}_\gamma)$ is the initial tuple with the maximal group size $n$ and corresponds to the target group. Then, the current group information is $G_{\pi^*}=\{\pi^*,id_{\pi^*},\ell^*,\Omega^*,st^*,\Delta^*\}$. A group index set $\mathbb{S}^*$ can be got from $st^*$ in $G_{\pi^*}$, where $\mathbb{S}^*=\{i|st^*[i]\neq \perp\}$. Based on $\mathbb{S}^*$, define $\mathbb{\bar{S}^*}=\{i|st^*[i]=\perp\}$ and $\mathbb{\bar{U}^*}=\mathbb{S}^* \setminus \mathbb{U}^*$. In this query, an abort event ${\sf Event\ 1}$ is defined. If the target group that $\mathcal{A}$ submits is not the $\pi$-th group or the target session round is not the $\ell$-th session round, we say ${\sf Event\ 1}$ happens. If ${\sf Event\ 1}$ doesn't happen, $\mathcal{C}$ does the following:  
	\begin{itemize}
		\item If $b=0$, compute $k_0=Z\hat{e}(g,Q)^{\sum_{i\in \mathbb{S}^*}a_i+\sum_{i\in (\mathbb{\bar{U}^*}\bigcup \mathbb{\bar{S}^*})}\alpha_{i\gamma}}$; otherwise, choose a session key $k_1$ from $\mathbb{G}_T$ at random.  
		\item Choose $b\in \{0,1\}$ randomly and return $k_b$ to $\mathcal{A}$. 
		
		
	\end{itemize}

	\medskip
	\noindent \textbf{Guess}: $\mathcal {A}$ submits $b' \in \{0,1\}$ to $\mathcal {C}$ as its answer. 
	
	We have known that $\mathcal {A}$'s advantage to win the above game is at least ${Adv}_{\mathcal{A}}$. To solve the decision $k$-BDHE problem, it requires $\mathcal{C}$ doesn't abort. That is, ${\sf Event\ 1}$ doesn't take place. It is easy to have $\Pr[\neg {\sf Event\ 1}]\geq \frac{1}{NL}$. Therefore, the advantage of $\mathcal {C}$ to solve the decision $k$-BDHE problem is at least $\frac{1}{NL}{Adv}_{\mathcal{A}}$. 
	
	$\hfill \Box$

	\section{Efficiency Evaluation}
	To evaluate the efficiency of our NI-CBE protocol, we first analyse the computational complexity of the protocol and then evaluate the performance of our protocol through simulations.

	\subsection{Complexity Analysis}
	Table \ref{Table1} presents the computational complexity of our NI-CBE protocol. In this table, the computation cost of {\sf GlobeSteup} algorithm is not analyzed since this algorithm only needs to be run once. That is, the efficiency of our protocol are mainly determined by the rest of algorithms. We note that some operations that can be pre-computed are not considered here.
	\begin{table}[htbp]
		\centering
		\caption{Computation Cost of the Algorithms}
		\label{Table1}
		\begin{threeparttable}
			\begin{tabular}{|c|c|}
				\hline
				Algorithms & Computation Cost \\ \hline
				KeyRegis &    $O(n)(T_E+T_M)$              \\ \hline
				KeyDerive &    $O(n^2+n)T_M+O(1)T_e$              \\ \hline
				KeyUpdate &   $O(1)(T_E+T_e)+O(n)T_M$       \\ \hline
				
				DCBEncrypt &    $O(s+s'+u')T_M+O(1)(T_e+T_E)$              \\ \hline
				DCBDecrypt &     $O(u')T_M+O(1)T_e$             \\ \hline
			\end{tabular}
			
		\end{threeparttable}
	\end{table}

	$T_E$/$T_M$ denotes the time to compute a scalar exponentiation operation/a scalar multiplication operation on the bilinear groups $\mathbb{G}_1$ and $\mathbb{G}_T$. $T_e$ denotes the time to complete a bilinear map operation. $n$ denotes the group size while $t$ represents the current number of group members of any group where a new party/old group member intends to join/leave this group. $s$ denotes the total number of existing group members in the target group before performing {\sf Encrypt} algorithm and $u$ represents the number of group members who are chosen as recipients within the target group. Then, we have $s'=n-s$ and $u'=s-u$. 
	
	\begin{figure}[]
		\centering
		\includegraphics[width=0.5\textwidth]{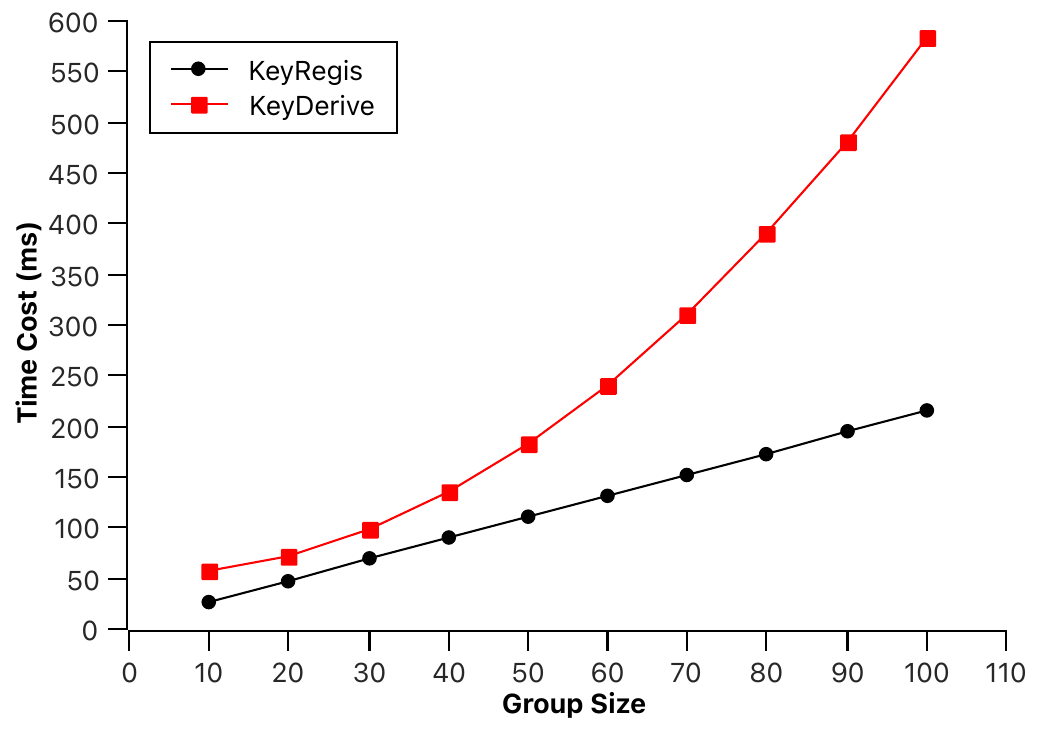}
		
		\caption{Time costs of {\sf KeyRegis} and {\sf KeyDerive}} \label{Fig 1}
		
	\end{figure}

	\subsection{Simulations}
	
	In this section, we simulated the running of the {\sf KeyRegis}, {\sf KeyUpdate}, {\sf Encrypt} and {\sf Decrypt} algorithm respectively. We note that the {\sf GlobeSetup} algorithm affects a little on the efficiency of the protocol since it is only invoked once. 
	The simulations were run on a Ubuntu machine with an Intel Core i7-4790 at a frequency of 3.6 GHz by using cryptographic library MIRACL. The security parameter was set to be 128 and a SSP curve with 128-bit security level was selected. The group size was set from $10$ to $100$, and the number of group members were set to be 80\% of each group size. The recipients were chosen from existing group members randomly every time running the {\sf Encrypt} algorithm. For simplicity, the operations that can be pre-computed were neglected in the simulations. 
	\begin{figure}[ht]
		\centering
		\includegraphics[width=0.5\textwidth]{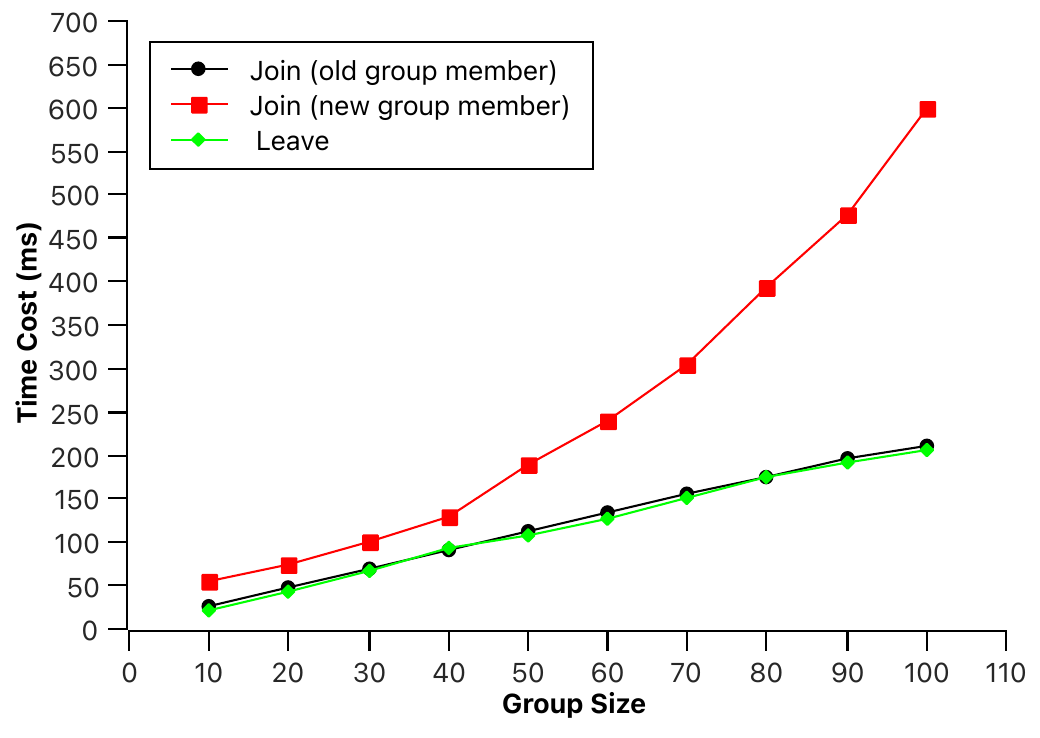}
		
		\caption{Time costs of {\sf KeyUpdate} } \label{Fig 2}
		
	\end{figure}

	Fig. \ref{Fig 1} presents the time costs of running {\sf KeyRegis} and {\sf KeyDerive}. It is easy to see that the running time of both algorithms scales with the group size. However, the group size has a more significant impact on the running time of {\sf KeyDerive}. When group size is $100$, the time costs of {\sf KeyRegis} and {\sf KeyDerive} are respectively less than $200$ ms and $600$ ms. Since {\sf KeyUpdate} consists of {\sf Join} and {\sf Leave} sub-algorithms, then we measured the running time of both of them. As shown in Fig. \ref{Fig 2}, for an old group member (existing in the group), the execution time of {\sf Join} increases linearly with group size. For a new group member wanting to join a group, the time cost of performing {\sf Join} grows with group size exponentially. One can see that the time cost of running {\sf Leave} approximately equals to that of running {\sf Join} for an old group member. When the group size is $100$, the overall execution time of {\sf Join}/{\sf Leave} is still acceptable (less than $200$ ms for an old member performing {\sf Join}/{\sf Leave} while less than $650$ ms for a new group member running {\sf Join}). Hence, the {\sf KeyUpdate} algorithm is efficient.
	

	\begin{figure}[]
		\centering
		\includegraphics[width=0.5\textwidth]{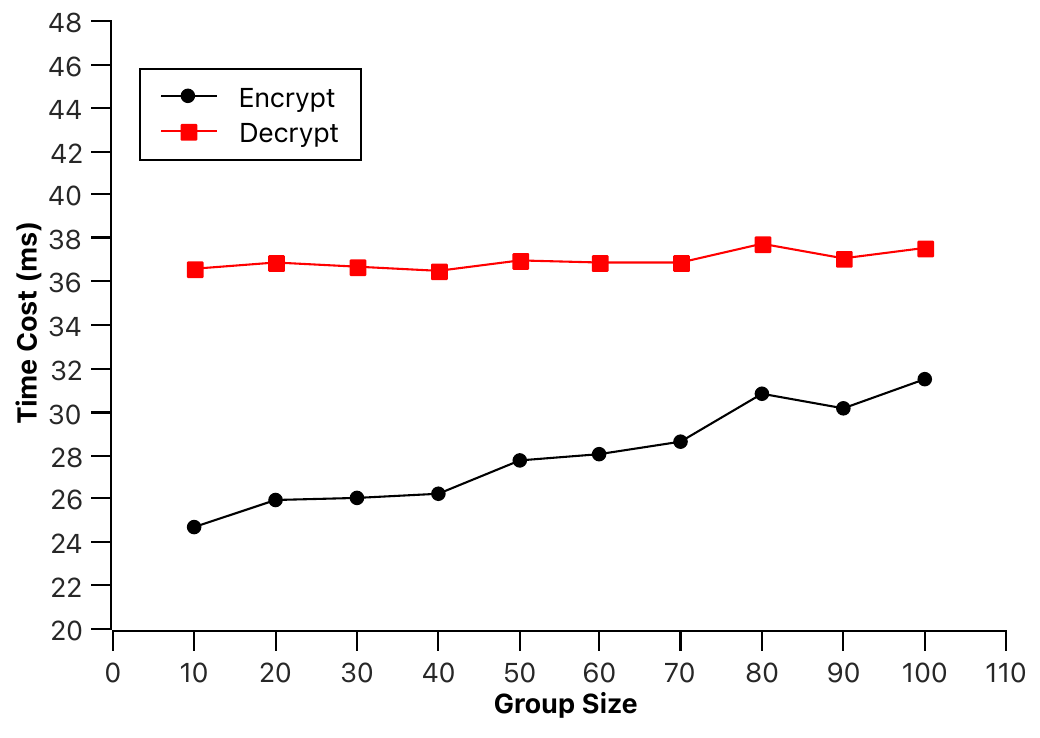}
		
		\caption{Time costs of {\sf Encrypt} and {\sf Decrypt}} \label{Fig 3}
		
	\end{figure}
	
	The time costs of running {\sf Encrypt} and {\sf Decrypt} are shown in Fig. \ref{Fig 3}. It is easy to see that the time cost of running {\sf Encrypt} grows slowly with the group size. This is because the execution time of {\sf Encrypt} is influenced by the number of recipients that increases correspondingly with the group size. Also, one can see that the time cost of running {\sf Decrypt} remains  constant for all group size. Overall, when the group size is $100$, the time cost for performing {\sf Encrypt} and {\sf Decrypt} is less than $32$ ms and $38$ ms respectively. Therefore, both {\sf Encrypt} and {\sf Decrypt} are efficient.

	\section{Conclusion}
We have proposed a non-interactive contributory broadcast encrytion (NI-CBE) protocol. This protocol is used by multiple parties who form a dynamic group to derive a public group encryption key and each party’s decryption key without requiring any interaction. Also, any party outside a group or any group member is allowed to join or leave the group still in a non-interactive way. 
More importantly, our protocol supports any party called a sender (even outside a group) to select some or all of group members and generate a ciphertext for them. This process still doesn't cause extra communication costs and the the size of ciphertext remains constant. We design a semi-statical security model to prove our protocol captures the correctness and indistinguishability of session key. Finally, we show our protocol is efficient through efficiency evaluation.

\bibliographystyle{plain}
\bibliography{main.bib}

\end{document}